\documentclass[12pt,preprint]{emulateapj}
\usepackage{apjfonts,psfig}

\shorttitle{Diagnostics for BHs in GCs}
\shortauthors{Noyola \& Baumgardt}

\begin{document}

\title{Testing Photometric Diagnostics for the Dynamical State and
  Possible IMBH presence in Globular Clusters}

\author{Eva Noyola} \affil{Max-Planck-Institut for Extraterrestrial
  Physics, Giessenbachstrasse 85748, Garching, Germany}
\affil{Universit\"ats-Sterwarte M\"unchen, Scheinerstrasse 1, 81679,
  Munich, Germany} 
\email{noyola@mpe.mpg.de}

\author{Holger Baumgardt}
\affil{Argelander Institute for Astronomy, University of Bonn, Auf dem H\"ugel 71, 53121, Bonn, Germany }
\affil{School of Mathematics and Physics, The University of Queensland, Brisbane, QLD 4072, Australia}
\email{h.baumgardt@uq.edu.au}

\begin{abstract}

  Surface photometry is a necessary tool to establish the dynamical
  state of stars clusters. We produce realistic \textit{HST}-like
  images from N-body models of star clusters with and without central
  intermediate-mass black holes (IMBHs) in order to measure their
  surface brightness profiles. The models contain $\sim$600,000
  individual stars, black holes of various masses between 0\% to 2\%
  of the total mass, and are evolved for a Hubble time. We measure
  surface brightness and star count profiles for every constructed
  image in order to test the effect of intermediate mass black holes
  on the central logarithmic slope, the core radius, and the
  half-light radius. We use these quantities to test diagnostic tools
  for the presence of central black holes using photometry. We find
  that the the only models that show central shallow cusps with
  logarithmic slopes between -0.1 and -0.4 are those containing
  central black holes. Thus, the central logarithmic slope seems to be
  a good way to choose clusters suspect of containing
  intermediate-mass black holes. Clusters with steep central cusps can
  definitely be ruled out to host an IMBH. The measured $r_c/r_h$
  ratio has similar values for clusters that have not undergone
  core-collapse, and those containing a central black hole. We notice
  that observed Galactic globular clusters have a larger span of
  values for central slope and $r_c/r_h$ than our modeled clusters,
  and suggest possible reasons that could account for this and
  contribute to improve future models.

\end{abstract}


\vspace{10pt}

\section{Introduction}\label{intro}

Surface photometry has often been the initial tool to establish the
dynamical state of globular clusters. The fact that the observed
radial density of most clusters appears to be well described by King
models \citep{kin66} has been taken as evidence that these clusters
are relaxed systems and that their dynamical evolution is dominated by
two-body relaxation processes. A natural consequence of two-body
relaxation is the onset of core collapse, where the central density of
a star cluster increases, while the core radius decreases (see section
1.2 of \citet{noy06} for a detailed description of the process). Some
clusters have been identified as having undergone core-collapse. These
are cases with very concentrated surface density profiles, showing
steep central cusps with a central projected logarithmic slope of
$\sim -0.7$ \citep{coh80}, that depart from King-type cores. About
20\% of the Galactic globular cluster population falls into this
category \citep{tra95}.

Kinematical evidence for core collapse accompanied by tailored models
have been presented for three clusters: M15 \citep{dul97}, NGC 6397
\citep{dru95}, and M71 \citep{dru92}. The expected velocity cusp has
only been resolved and modeled for M15
\citep{bau03a,mcn04,bos06}. NGC~6752 has been considered to be a post
core-collapse cluster by many authors, but different datasets and
analysis methods find that it has a small flat central core
\citep{lug95,fer03,noy06}. The size of the core might be consistent
with models for gravothermal oscillations \citep{ves94}. This cluster
shows a steep central velocity cusp \citep{dru03}, but no tailored
core-collapse model has been created for it.

There are a variety of heating mechanisms that can drive energy into
the core of a star cluster, causing it to expand, and thus preventing
core-collapse. The effect of binary heating by primordial binaries is
the best studied mechanism to date \citep{gao91,ves94}, although it
has been proposed that most Galactic globular clusters are not yet in
the binary-burning phase of evolution \citep{fre08}. The presence of
stellar-mass black holes acting as an energy source, has recently been
invoked to explain the distribution of core sizes in LMC and SMC
globular clusters \citep{mac08}. Mass loss by stellar winds during
early times of the cluster evolution also contributes to cluster
expansion \citep{bau07,hur07,bas08}, so clusters might expand
considerably even if they are born with concentrated configurations. A
recently proposed mechanism is velocity kicks imparted during white
dwarf formation, which would also act as a heating mechanism for
clusters with velocity dispersions of a few km/s \citep{dav08,fre09}.

The presence of a central intermediate mass black hole (IMBH) of
100-10,000 M$_\odot$ is another mechanism that can affect the
dynamical evolution of star clusters. \citet{bah76} calculated the
shape of the radial density profile for a single mass star cluster
around a massive black hole. They predicted the formation of a steep
central cusp with a logarithmic slope of -1.75. \citet{bau04a,bau04b}
confirmed these results based on direct N-body simulations. They also
showed that multi-mass clusters with IMBHs are mass-segregated in
their centers and that main-sequence stars have cusps that are
significantly flatter than $-$1.75. They found that the IMBHs appear
to produce shallow central cusps on the projected density profiles of
bright main sequence stars for these clusters, with slopes of $\sim
-$0.2, as opposed to steep power-laws \citep{bau05}. \citet{noy06} and
\citet{noy07} (hereafter called respectively NG06 and NG07) obtained
surface brightness profiles from \textit{HST} images for Galactic,
LMC, SMC and Fornax dwarf galaxy globular clusters. They found that
about 20\% of the globular clusters in their sample show central
slopes in this intermediate range.

The surface density profile shape can also be affected in the size of
its core when a central black hole is present.\citet{tre07} estimated
the value of $r_c/r_h$ (ratio of core radius to half-light radius) for
N-body simulated star clusters containing central black holes. They
used a density averaged radius as a measure for the core radius. They
found that the ratio tends to reach values around 0.3 for these cases,
while the value is considerably smaller ($< 0.1$) for clusters without
black holes. On the other hand, \citet{hur07} finds similarly large
$r_c/r_h$ values for N-body simulations evolved including $0.5-10$\%
primordial binaries, but without a central black hole. In this case,
the Casertano \& Hut method \citep{cas85} was used to obtain the
three-dimensional core radius. The different way in which the core
radius was measured from the N-body simulations differs between the
  two results. Recently, \citet{ves10} analyzed direct N-body models
with and without IMBHs. They found that shallow cusps with logarithmic
slopes as steep as $-$0.3 are present in various models, not only the
ones containing black holes. The apparent discrepancies between
different models using different analysis techniques stresses the
importance of performing meaningful measurements on models so they can
be properly compared with observational data.

Direct dynamical evidence for the existence of central black holes
using velocity dispersion measurements has been put forward for three
nearby globular clusters. M15 was the first case \citep{ger02,ger03},
but alternative models without black holes were also shown to be good
fits to the data \citep{bau03a}. The latest detailed dynamical
measurement and model finds non-conclusive evidence for the presence
of a central black hole in this cluster \citep{bos06}. G1, a large
globular cluster in Andromeda, has stronger observational evidence to
support the presence of a central black hole, from integrated
kinematical measurements \citep{geb03,geb05}, as well as from X-ray
\citep{poo06} and radio \citep{ulv07} observations, but alternative
scenarios have also been presented for this case \citep{bau03c}. Omega
Centauri is the most recent case for which line-of-sight velocity
dispersion measurements appear to support the existence of a central
black hole of 40,000 $M_\odot$ \citep{noy08}, but proper motion
measurements from HST images find different results
\citep{and10,vdm10}. Evidence has also surfaced for intermediate mass
black holes in extra-galactic disk galaxies based on X-ray
observations.

Ultra luminous X-ray sources (ULXs) have X-ray luminosities higher
than the Eddington limit for a stellar mass black hole. One of the
possible explanations for this emission is that it comes from
accretion onto an IMBH. For example, the galaxy M82 contains an ULX
source which is believed to host an IMBH based on the absolute
brightness of the source \citep{mat99,mat01}, and its radio
variability \citep{str03}. The position of the X-ray source appears to
coincide with the young dense star cluster MGG-11 \citep{mcc03}. There
is also the controversial case of the globular cluster RZ 2109 in
NGC~4472, which shows the first clear evidence for a star cluster
hosting a black hole \citep{mac07}, but the size of the black hole is
still under debate \citep{zep08}. One more interesting object is the
X-ray source CXOJ033831.8-352604, associated with a globular cluster
in the Fornax elliptical galaxy NGC 1399. \citet{irw10} suggest the
emission might come from a tidally disrupted white dwarf around an
IMBH.
 
In this paper, we create synthetic HST-like images from N-body
simulations with and without IMBHs. We measure their surface
brightness profiles as we would with observations. We provide an
analysis of the detailed shape of central density profiles for these
models that helps understand the central state of Galactic globular
clusters. We describe the N-body models in section 2, the synthetic
images in section 3, data analysis in section 4, and discussion in
section 5.

\begin{deluxetable}{lccccccc}
\tablewidth{0pt}
\tabletypesize{\huge}
\tablecaption{\label{tab1}N-body models.}
\tablehead{
\colhead{model} &
\colhead{source}&
\colhead{T$_e$}   &
\colhead{M$_{\bullet}$/M$_{TOT}$} &
\colhead{W$_0$} &
\colhead{total N}  &
\colhead{input N}   &\\
\colhead{} &
\colhead{} &
\colhead{Gyr}   &
\colhead{$M_\odot$} &
\colhead{} &
\colhead{$10^3$ stars}   &
\colhead{$10^3$ stars}   &
}
\startdata
m1t1.0  & BM03 &  1.0 & \nodata & 7  & 515 & 342 \\
m1t4.0  & BM03 &  4.0 & \nodata & 7  & 474 & 308 \\
m1t7.0  & BM03 &  7.0 & \nodata & 7  & 444 & 298 \\
m1t9.0  & BM03 &  9.0 & \nodata & 7  & 439 & 294 \\
m1t10.0  & BM03 & 10.0 & \nodata & 7  & 401 & 271 \\
m1t11.0  & BM03 & 11.0 & \nodata & 7  & 416 & 282 \\
m1t12.5  & BM03 & 12.5 & \nodata & 7  & 394 & 267 \\
m1t16.0  & BM03 & 16.0 & \nodata & 7  & 354 & 241 \\
m2t2.0  & BM03 &  2.0 & \nodata & 5  & 513 & 219 \\
m2t6.0  & BM03 &  6.0 & \nodata & 5  & 414 & 220 \\
m2t8.0  & BM03 &  8.0 & \nodata & 5  & 361 & 221 \\
m3t2.0  & \nodata &  2.0 & \nodata & 5  & 507 & 163 \\
m3t5.0  & \nodata &  5.0 & \nodata & 5  & 587 & 165 \\
m3t8.0  & \nodata &  8.0 & \nodata & 5  & 529 & 141 \\
m3t11.0  & \nodata &  11.0 & \nodata & 5  & 560 & 148 \\
m3t14.0  & \nodata &  14.0 & \nodata & 5  & 567 & 151 \\
 & & & & \\
\cline{1-7}
 & & & & \\
mb1t11.5 & BMH05   & 11.5 & 0.2\% & 7 & 517 & 239 \\
mb2t11.0 & BMH05   & 11.0 & 0.5\% & 7 & 519 & 240 \\
mb2t11.5 & BMH05   & 11.5 & 0.5\% & 7 & 517 & 236 \\
mb2t12.0 & BMH05   & 12.0 & 0.5\% & 7 & 516 & 233 \\
mb3t11.5 & BMH05   & 11.5 & 1.0\% & 7 & 515 & 223 \\
mb4t3.0 & BMH05 & 3.0 & 2.0\% & 7 & 520 & 188 \\
mb4t6.0 & BMH05 & 6.0 & 2.0\% & 7 & 519 & 159 \\
mb4t9.0 & BMH05 & 9.0 & 2.0\% & 7 & 518 & 138 \\
mb4t11.0 & BMH05   & 11.0 & 2.0\% & 7 & 518 & 224 \\
mb4t11.5 & BMH05   & 11.5 & 2.0\% & 7 & 516 & 220 \\
mb4t12.0 & BMH05   & 12.0 & 2.0\% & 7 & 515 & 216 \\
mb5t11.3 & \nodata & 11.3 & 2.0\% & 7 & 516 & 186 \\
mb5t11.8 & \nodata & 11.8 & 2.0\% & 7 & 516 & 184 \\
mb5t12.0& \nodata & 12.0 & 2.0\% & 7 & 515 & 180 \\
\enddata
\tabletypesize{\normalsize}
\end{deluxetable}
\vspace{0cm}

\vspace{0pt}

\section{N-body models}\label{models}

It is often challenging to make a direct comparison between the
results of N-body models and observations because it is hard to take
into account the sources of uncertainty of observations. It is
complicated to discriminate how much of the information from the
models would be available to an observer if the simulated object was
on the sky at a realistic distance. With the goal of making more
meaningful comparisons, we take the output of N-body models and create
realistic synthetic images from them.

The simulations used here followed the evolution of star clusters with
and without central intermediate-mass black holes. All star clusters
contained 131,072 (128K) stars initially and were simulated with the
\mbox{N-body} program NBODY4 \citep{aar99} using the GRAPE-6 computers
at Tokyo University.  Stellar evolution was followed using the fitting
formula of \citet{hur00}, assuming a metallicity of Z=0.001. For most
models, the initial density profile was given by a King $W_0=7$
configuration but we also include two models that started from a King
$W_0=5$ configuration. The detailed description of the runs can be
found in \citet{bau03b} and \citet{bau05}, called BM03 and BMH05
respectively.

Models m1t and m2t are models without IMBHs. The data was taken from
the N=128K star runs in BM03b, who assumed a neutron star retention
fraction of 10\% and a mass range between 0.1 and 15~$M_\odot$,
according to a \citet{kro01} mass function. Model m3t is a new model
made for this paper, starting with N=128K stars distributed according
to a \citet{kro01} IMF from 0.1 to 100 $M_\odot$, and with an asuumed
neutron star and stellar black hole retention fraction of 10\%
\citep{pfa02a,pfa02b}. The retention fractions are assumed to be the
same for simplicity, since there is still considerable uncertainty
about these numbers. This simulation contains stellar mass black holes
with masses up to 25 $M_\odot$ until the the stellar-mass black holes
have kicked each other out in two and three-body interactions at about
T=12Gyr. All models without black holes are at a galactocentric
distance of 8.5 kpc. Models mb1t to mb4t come from BMH05, in this case
the neutron star retention fraction was 15\%, and the stellar mass
range went from 0.1 to 30~$M_\odot$, assuming the same mass function
as for the non IMBH cases. The models contain IMBHs of masses 0.2\%,
0.5\%, 1.0\%, and 2.0\% of the total mass of the star cluster
(M$_{TOT}$). If we extrapolate the scaling laws for super-massive
black holes in galaxies to the mass regime of globular clusters, the
case with $M_\bullet$=0.5\% $M_{TOT}$ would follow the Magorrian
relation \citep{mag98}. Stars passing close to the IMBH were assumed
to be tidally disrupted. We use the \citet{koc92} formula for the
disruption radius. We also performed one additional simulation of a
star cluster with an IMBH (called mb5t in Table~1). For this
simulation we overlayed four snapshots of the mb4t cluster at $T=11$
Gyr and continued the simulation for 1 Gyr with $N=508.000$
stars. Given the large number of stars, no stacking was necessary for
this cluster. We use this model to test if there is any effect from
the stacking of close snapshots.

For the non IMBH models, we created different snapshots in order to
investigate the core-collapse evolution. Snapshots were taken at 1.0,
4.0, 7.0, 9.0, 10.0, 11.0, 12.5, and 16.0 Gyrs for the m1t case; and
2.0, 6.0, and 8.0 Gyrs for the m2t case. For the m1t model,
core-collapse occurs at 12.5 Gyrs, while this happens at T=21.3 Gyrs
for the m2t model and at T=20.5 for model m3t. For the models
containing IMBHs we use snapshots at different evolutionary times, all
between 11 and 12 Gyr, except model mb4t, for which we have earlier
snapshots. Information extracted from the models include mass,
position, V magnitude, and temperature of each star. The details of
the created models can be found in Table~\ref{tab1}.

\begin{figure}[t]
  \centerline{\psfig{file=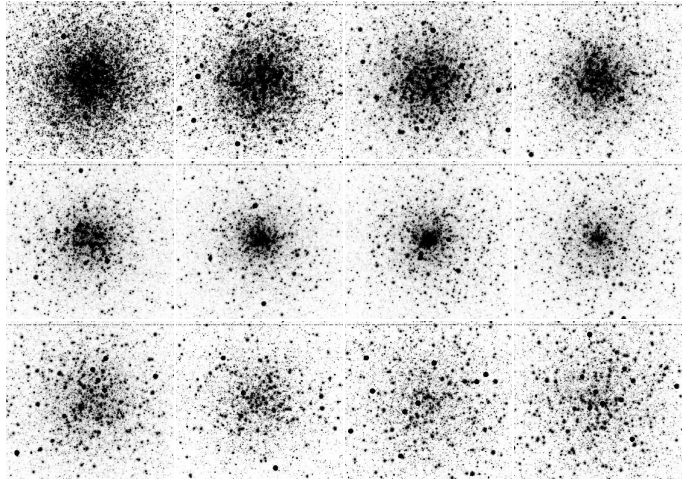,width=9.0cm,angle=0}}
  \figcaption{Synthetic images for various N-body simulations. The top
    and middle rows show a cluster without a central IMBH at different
    evolutionary times of 1.0,4.0,7.0,9.0,10.0,11.0,12.5, and 16.0 Gyr
    (models m1t1.0-m1t16.0 of Table~1). The bottom row shows clusters
    containing IMBHs of 0.2\%, 0.5\%, 1.0\% and 2.0\% $M_{TOT}$
    (models mb1t11.5,mb2t11.5,mb3t11.5, and mb4t11.5 of Table~1).The
    evolution toward core-collapse for the non IMBH case is clearly
    visible on the images in the upper and middle rows.}
\label{im1}
\end{figure}

\begin{figure*}[t]
\centerline{\psfig{file=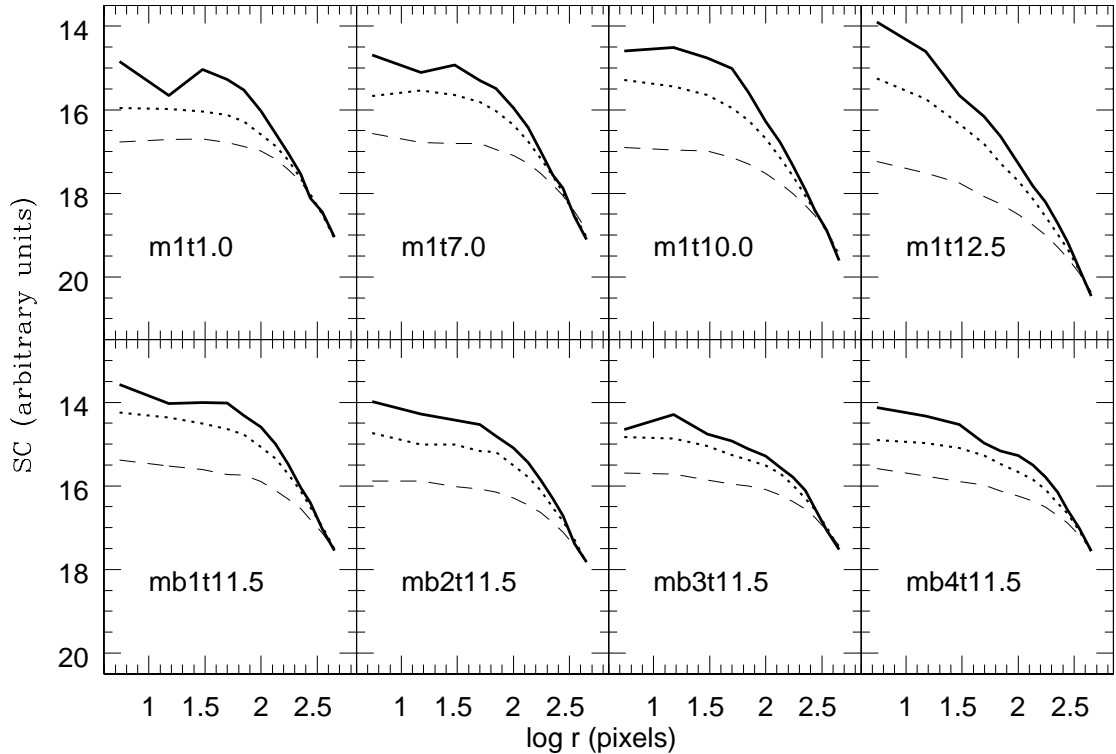,width=16.0cm,angle=0}}
\caption{Star count projected density profiles from the input lists,
  which contain more stars than those detected in the synthetic
  images. For every case, the magnitude bins are: V$_{mag}< 16$ (solid
  line), $16<$V$_{mag}<20$ (dotted line), and V$_{mag}> 20$ (dashed
  line).The top row shows a case without an IMBH at evolutionary times
  of 1.0, 7.0, 10.0 and 12.5 Gyrs (models m1t1.0, m1t7.0, m1t10.0, and
  m1t12.5). The bottom row shows cases containing IMBHs of 0.2\%,
  0.5\%, 1.0\% and 2.0\% $M_{TOT} $ (models mb1t11.5, mb2t11.5,
  mb3t11.5, and mb4t11.5). The evolution towards core-collapse affects
  the bright and intermediate bins, but not the faintest one. The
  presence of an IMBH affects the central slope of brighter bins, and
  the core radius of every stellar group.}
\label{im2}
\end{figure*}

Given the initial number of stars and their mass function, stellar
evolution, tidal evaporation, and disruption of stars by the IMBH, the
final mass of the models is between $15,000M_\odot$ to
$45,000M_\odot$, which is only about 1-10\% the mass of a typical
Milky Way globular cluster.  Since the analysis performed in this work
requires a large signal in the images, we had to resort to stacking
snapshots separated by short periods of time around a given age for
every model. For the models without an IMBH, we stack $\sim$10
snapshots separated by 15 Myrs, while for the models containing IMBHs
we stack 5 snapshots separated by 5Myrs. The ultimate goal is to have
the same number of stars in the central region for every model. Models
with IMBHs are not subject to any external tidal forces, while the
models without IMBHs are placed on a circular orbit around a Galactic
tidal field and therefore undergo a stronger mass loss. In the end,
the total number of stars present in our original lists is always
around $N\sim500,000$. The total mass for the stacked models is $\sim$
220,000 $M_{odot}$ for the non-BH models and $\sim$ 180,000 $M_{odot}$
for models with central BHs. Variations between individual models are
under 10\%.

The total number of stars for each N-body model is given in column~6
of Table~\ref{tab1}. The number of stars included in the synthetic
images is given in column~7. As explained in detail in
Section~\ref{images}, this constitutes only $\sim50\%$ of the original
list due to brightness and radius cuts. The modeled clusters are also
more extended compared to the Galactic clusters. With the goal of
making the modeled clusters look more like dense Milky Way clusters,
we also scaled the clusters down in size. We do this by dividing their
coordinates by a common factor, which we chose to be a $\sim8$ for the
models without IMBHs, and a factor of 4 for the models with
IMBHs. This scales all clusters down to a similar half-light radius
(7-10 pc for non-collapsed cases), which is similar to that measured
for Galactic globular clusters.

\vspace{0pt}

\section{Creating synthetic images}\label{images}

Our main goal is to create realistic images from the N-body models in
order to perform the same type of analysis that we do on \textit{HST}
observations. The quality and size of the images is chosen to match
that of the PC chip in WFPC2 or the HRC channel in ACS. In this way,
we can make a proper comparison with observed clusters contained on
NG06.

The procedure to create images is like the one described in detail in
NG06 and NG07. We use DAOPHOT \citep{ste87} to add stars from a list
of positions and magnitudes onto a base image. With the goal of
including realistic background noise, we use as a base a WFPC2 image
of a sparse field with the few present stars cleanly subtracted. We
modify the base image to have a larger number of pixels than the PC
chip on WFPC2, and we locate the center of the cluster at the center
of the base image. The utilized point spread function (PSF) is
obtained from observed data and it does not include variations across
the chip.

Since the center of observed clusters is not known a priori, we made a
blind test, in which the center of the models were given an arbitrary
shift in the three spatial coordinates, and the new center was
calculated using the octants method described in detail on NG06. We
choose a guess center and a radius, we count the stars present in
eight 'pie slices' segments defined by the chosen center and radius
and we calculate the standard deviation of the eight numbers. Using
the same radius, we move to a new guess center and repeat the
procedure several times around the initial guess center. In the end we
have a map of center locations and a standard deviation value
associated to each of them. We fit a smoothing spline to the resulting
surface and find the location of the minimum defined by the grid of
guess centers, which we take as the true center. For this procedure,
we used every star in the list, which implies using many more stars
than the ones that would be available to an observer. Our goal is to
test the method for a complete dataset, not to test the observed
accuracy of the measurement, since this has already been tested in
NG06 and NG07. The centers were calculated for three projections on
the x-y, x-z, and y-z planes. In the end we found that the method is
able to recover the center with an accuracy of 0.01pc
($\sim$0.005$r_c$). The tests performed in NG06 yield an error for the
observed center location that corresponds to $\sim$0.05$r_c$.  In
general, the effect of measuring a density profile using the wrong
radius is not necessarily to change the central surface brightness
slope, but instead, a drop in the central measurement point is
created. Seeing such a drop is actually an indication of having the
wrong center, \citet{lan08} use this fact as a test for correct
centering in their work, for example. Despite that drop, the slope of
the other points up to the core radius is normally the same as the one
using the correct radius. This is clearly seen comparing the profiles
for omega Centauri between \citet{noy08} and \citet{and10}. Despite
using very different centers, and getting density profiles with
different shapes, the slope of the profile between 15" and the core
radius is consistent in both cases.

\begin{figure*}
\centerline{\psfig{file=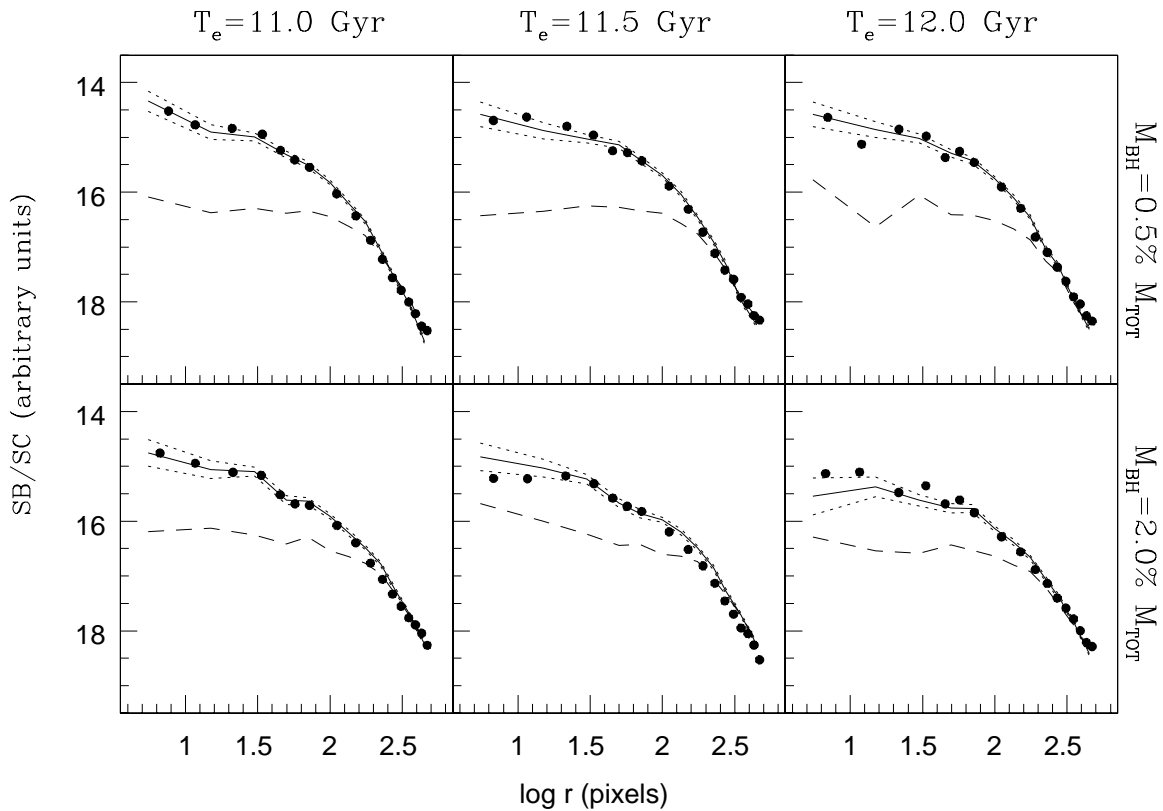,width=16.0cm,angle=0}}
\caption{Surface brightness and star count profiles for models with
  IMBHs of 0.5\% and 2.0\% $M_{TOT}$ at evolution times of 11.0, 11.5,
  and 12.0 Gyrs. The solid line is the N-body profile, the dotted
  lines mark the Poisson error for the N-body profile, the dashed line
  is the star count profile, and the filled points are the measured
  photometric points. The vertical scale is arbitrary. Uncorrected
  star counts underestimate central surface densities by a factor of
  2-5, while photometric measurements are better suited to determine
  density profile for crowded fields, they always lie within the
  errors of the N-body profile.}
\label{profbh}
\end{figure*}

The next step for making star lists suitable to be turned into images
is projecting the stellar coordinates into a 2D distribution on the sky. We
need to assume a fiducial distance that will affect both the
coordinates and the magnitude of each star. The chosen distance for
all cases was 5 kpc, which is on the near end of the distribution of
distances for Galactic clusters. We choose this distance since it is
adequate for our goal of obtaining high signal to noise images. After
performing the geometrical projection and applying the distance
modulus to the star's magnitudes, bolometric corrections are performed
to obtain the V-band luminosity of each star. The correction is done
taking into account the star's temperature following the procedure by
\citet{hur00}. We create synthetic images using DAOPHOT, which has a
fiducial zero photometric point of 25 magnitudes, therefore, we
eliminate from our list all stars fainter than that. These faint stars
constitute $\sim20\%$ of the entire list. At 5 kpc distance, 1pc
radius is equivalent to 41.25 arcsec. Assuming a pixel scale of
0.1\arcsec per pixel, this is equivalent to $\sim$412 pixels. Taking
into account the extra scaling factor mentioned in
section~\ref{models}, the synthetic images (1000 pixels on the side)
contain stars inside a radius of $\sim$10 pc for each simulated
cluster. Since we are interested in the central structure of the
clusters, we choose the image size to include approximately 10
core-radii, and we exclude stars outside this radius. The final images
end up including $\sim50$\% of the total number of stars in the
simulated clusters.

The results for a subset of the models can be seen in Fig.~1. For the
model without an IMBH, it is clear that the cluster achieves a very
concentrated configuration as it evolves towards core-collapse. On the
other hand, the clusters containing central IMBHs are less dense and
have more extended cores. Once we have the synthetic images, we
proceed to analyze them in the same way as we do with observed data.

We count the number of detected stars inside the average core radius
for our models. The average detected stellar density in this region is
$\sim$2 stars/arcsec$^2$. For comparison, this is an order of
magnitude lower than the central density detected for NGC~6388 by
\citep{lan07}. From the \citet{noy06} compilation, we located two
clusters at different heliocentric distaces that have a similar
central densities to our models, NGC~5634 (30 Kpc) and NGC~6541 (7.5
Kpc) \citep{har96}.

\section{Surface Density Profiles}\label{sb}

We measure surface density profiles for every synthetic image
following the prescription described in detail in NG06. Using various
DAOPHOT routines, we find stars and then perform PSF-fitting
photometry on them. DAOPHOT allows for the inclusion of noise when
adding synthetic stars, therefore, even when we utilize the same PSF
used to create the images for our photometric measurements, the
subtractions are not perfect and are comparable to those in observed
data. We have tested our measurement methods thoroughly using
simulated images in NG06 and NG07. We know that we can measure the
input centers within $\sim$1 \arcsec for concentrated clusters,
therefore, we directly use the known input center for every image when
we measure density profiles.

\begin{figure*}
\centerline{\psfig{file=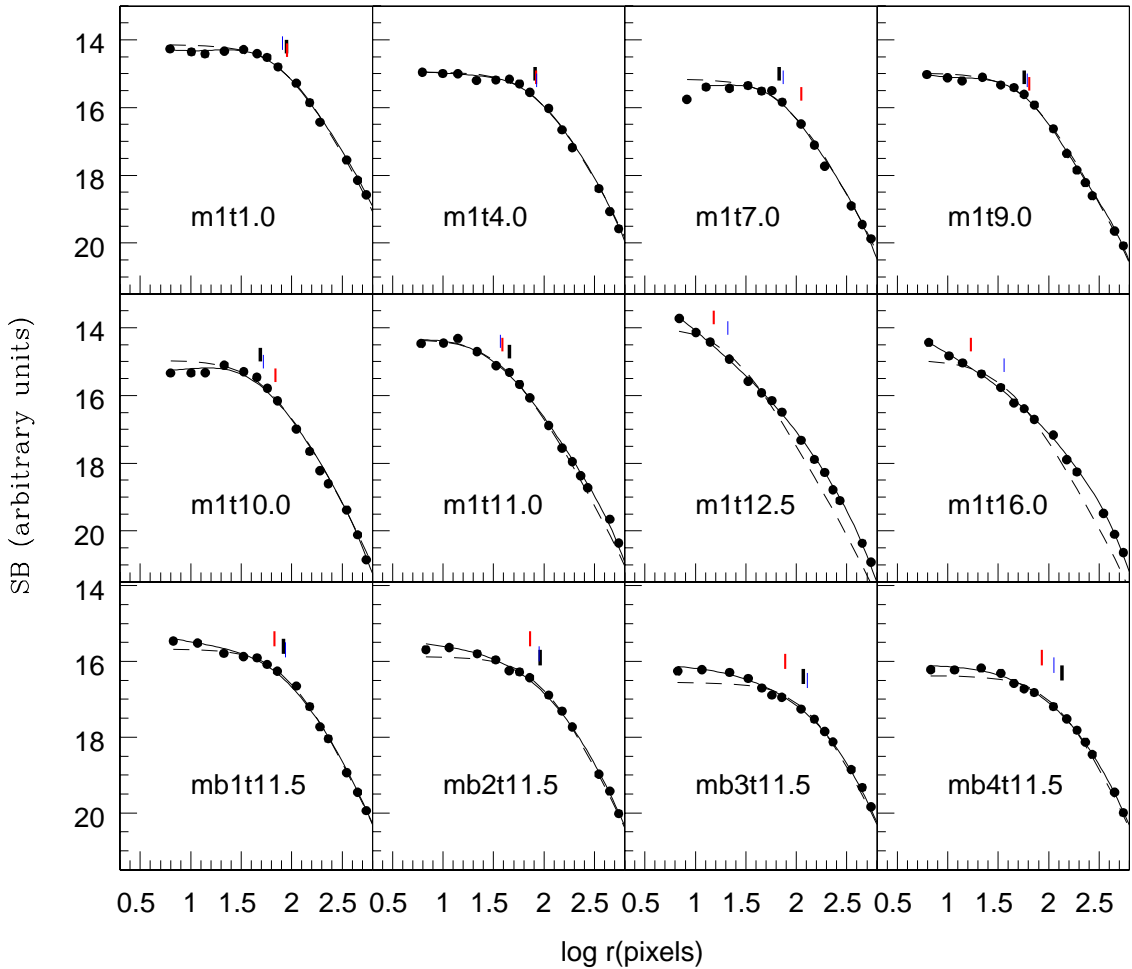,width=16.0cm,angle=0}}
\caption{Photometric points for various snapshots of models with and
  without IMBHs. The solid line is a smooth profile from the data
  points, the dashed line is a single-mass King fit. The vertical
  lines mark the different measured radii: black (thickest) is break
  radius, blue marks radius from King fit, and red (thinnest) marks
  FWHM core radius.}
\label{profnobh}
\end{figure*}

The density profiles are obtained in two different ways: from
integrated light and using star counts. A detailed discussion of the
pros and cons for each method can be found in section 2.3 of NG06. For
the first method, we use the magnitudes of detected stars to identify
the brightest 2-3\%, and we then proceed to mask them by giving them a
value that excludes them from the integrated light measurement. For
most stars, we assign a masking radius of 3 pixels, which only
eliminates the central bright region of stars, not the
halo. Occasionally, if very bright stars are present near the center
of the cluster, we use a larger radius to mask those. The haloes of
the stars do contribute to the total light, but by masking the central
part of the PSF disk, one prevents the giant stars from dominating the
measurements. Obviously, the "contamination" effect is stronger in the
very central regions in cases where there are many giant stars (like
in post core-collapse). In our models, the integrated light follows
the input profile very closely even in these cases.  Also,
\citet{lut11} perform detailed PSF contribution estimations for ACS
imaging of NGC 6388. They conclude that the contribution of bright
stars after masking the central part of the PSF is under 10\% for bins
containing 10 pixels or more. Our bins are always larger than that.

The number of detected stars is roughly 10\% of the input stars,
although it is worth pointing out that about 70\% of the input stars
are fainter than 20th magnitude. These stars make an important
contribution to background light, but they are only detected as
individual sources with low efficiency. As expected, the detection
efficiency is close to 100\% for the brightest stars (V$_{mag}$<16),
while the percentage declines for fainter stars, particularly closer
to the center where crowding problems are worse. Once we have masked
the 3\% brightest stars, we measure integrated light by calculating
the number of counts per pixel in various annuli using the biweight, a
statistically robust estimator \citep{bee90}. As discussed in detail
in NG06, this appears to be the optimal way to extract a density
profile for stars with mass at or around the turnoff mass for an
evolved cluster. The choice of the sizes for the annuli is a tradeoff
between obtaining the highest spatial resolution and obtaining the
least noisy profile possible.

The second method we use to measure density profiles is star counts.
From a star list, we construct a star count profile in the same annuli
where we measure integrated light. This is done by estimating the
number of stars per unit area, where every star has the same
weight. As mentioned above, it is well known that in crowded field
photometry, fainter stars are detected with decreased efficiency.  The
exact completeness fraction for a given brightness at a given radius,
depends on the specific shape of each profile.  Given that the surface
brightness profiles are dominated by the brightest stars, we measure
star count profiles only for the stars brighter than a given magnitude
for each cluster, since this is the only way to make a meaningful
comparison between the two methods. In order to obtain formally
correct star count profiles from images, one must calculate the
correct completeness correction factor for each brightness group in
each image, which is very time consuming and outside the scope of this
work. Uncorrected star counts have been used to measure density
profiles for star clusters recently (e.g. Lanzoni et al., 2008), so we
feel that it is relevant to compare to such profiles. For the models
containing IMBHs, the brightness cutoff always corresponds to 16
magnitudes (slightly fainter than the turn-off point, equivalent to
stars with 0.8 $M_\odot$); for the non IMBH models, the limiting
magnitude changes with evolution time and is brighter than 16
magnitudes for every case except the most evolved case at 16 Gyrs. We
use these limiting magnitudes to calculate a star count profile from
the original input list, as opposed to the detected list, and we call
this the 'N-body profile'. We compare our measured profiles against
this N-body profile, which can be thought of as the 'true' profile of
the cluster, since it comes straight out of the entire model
dataset. The precise limiting magnitude for each model is taken as the
one for which the surface brightness profile matches the N-body
profile in the region outside the core radius.

We notice that every simulated cluster, with and without IMBHs, shows
mass segregation, as can be seen in Fig.~\ref{im2} where we compare
profiles obtained from the input list for various brightness
groups. As expected, the profiles for the brightest stars are more
concentrated than for the intermediate and faintest groups. The
faintest group almost always shows flat central densities, except for
the case containing a 2.0\% M$_{TOT}$ IMBH. As explained above, the
solid lines in this figure are taken as our 'N-body' profile for every
model.

We compare the measured surface brightness and star count profiles
with the N-body profiles. This is shown in Figure \ref{profbh} where
we present three profiles for models with IMBHs of 0.5\% and 2.0\%
$M_{TOT}$ at different evolution times of 11.0, 11.5, and
12.0~Gyrs. The limiting magnitude for the N-body profile and the
measured star count profile is always the same. We note that the
uncorrected star count profiles always underestimate the density for
the central regions, including at and around the core radius, while
all three profiles agree very well at large radii. The N-body profile
is sometimes noisy at the center, which is expected due to the small
numbers of bright stars in that region. We show the Poisson noise for
the N-body profile. As can be seen, for every case, the integrated
light profile follows the N-body profiles very well at $r>30$\arcsec,
and is as smooth as the N-body profile inside the core. The shape of
the surface brightness profile is clearly dominated by the brightest
stars, but the masking of the bright stars combined with the
background contribution from fainter stars helps to make it smooth.
It practically always lies within the Poisson errors for the N-body
profile.

Once we have obtained the photometric points for each case, we use a
smoothing spline \citep{wah90} in order to obtain a smooth profile for
further analysis. Since we want to measure half-light radii as well as
fit King profiles, we need to cover the complete radial extent for the
clusters. Given that both surface brightness and star counts agree
very well with the N-body profile at large radii, we extend the
measurements using the N-body profile to the complete radial extent of
each modeled cluster. We decide to truncate the star counts at the 0.2
pc width annulus for which we no longer detect stars. There might be
stars present at larger radius, but we know that they are very few. The
lower density limit we use is lower than what one could measure for observed
Galactic clusters, where the field population already would dominate
the measurements. In the end, we fit a smooth spline to a combination
of our measured photometric points for the radial extent of our
images, and of the N-body profile at larger radii.

\begin{figure}[t]
  \centerline{\psfig{file=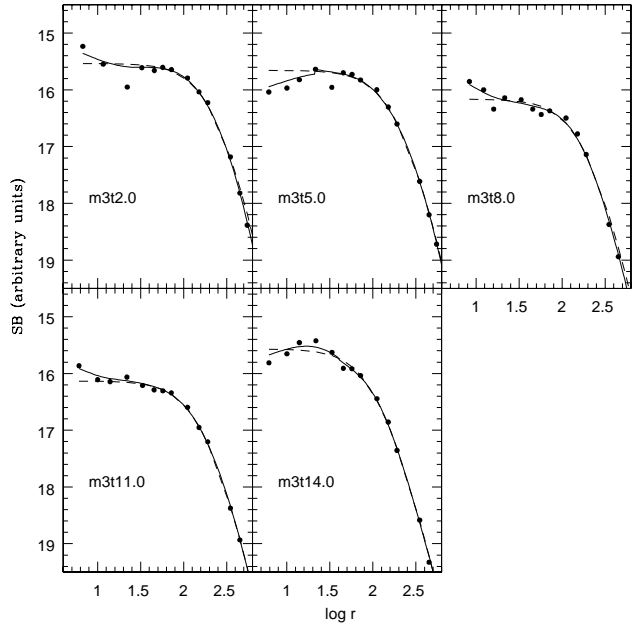,width=9.0cm,angle=0}}
  \figcaption{Photometric points for various snapshots of models
    without IMBHs but containing stellar-mass black holes. The solid
    line is a smooth profile from the data points, the dashed line is
    a single-mass King fit.}
\label{radslope}
\end{figure}

\section{Analysis}\label{analysis}

As mentioned in Section \ref{intro}, two types of photometrical
measurements have been proposed as possible diagnostics for the
presence of IMBHs in star clusters, the central slope of the density
profile, and the $r_c/r_h$ ratio. In this section we explain how we
obtain both quantities for our simulated clusters.

The measurement of the half-light radius ($r_h$) is straightforward
once we have the complete smooth profiles. We integrate the light
profile to get the total luminosity and take the radius at which the
profile contains half the amount of light. The measurement of the core
radius is more complicated since there are different definitions and
ways to measure it for observations and numerical modeling. In this
work, we explore three different ways to measure core radii that are
normally used for observed clusters. The first is the one used by
\citet{tra95} and \citet{har96}, whose results are the sources for
most studies of large samples of Galactic globular clusters. These
catalogs define the core radius as the half width half maximum of the
radial density profile ($r_{ch}$ from now on). This definition makes
the radius resolution dependent when the profiles are not flat towards
the center, since the closer to the center we measure, the brighter
the central luminosity value becomes. A second definition is the one
that comes from fitting a single-mass King profile \citep{kin62} to
the density profile and taking the value of the fit for the core
radius (which we call $r_{ck}$). The third definition is the one used
in NG06, called break radius, and defined as the radius of maximum
curvature of the density profile (called $r_b$). It can be understood
as the turnover radius.

As can be seen in Figure~\ref{profnobh} and Table~\ref{tab2}, the
agreement between the three radii is good for models with central
slopes between 0.05 and -0.05 (i.e., those with flat central
cores). As expected, $r_{ch}$ is smaller than the other two radii for
models with central cusps. It can also be seen that the King fits
agree very well with the observed profiles for models with flat cores,
while for the rest, the agreement of the King fit is good only outside
the core radius, but the values of $r_{ck}$ and $r_b$ start to
diverge. Notice that the King fit does not describe the profile well
towards the center. For models m1t12.5 and m1t16.0, the profiles are
so steep, that we cannot measure a reliable minimum of curvature
($r_b$). The King fit for these cases is a bad match for the entire
radial extent, so, even though one can formally obtain a value for
$r_{ck}$ and $r_{ch}$, neither of them provide meaningful information
about the density profile. For the cases in which we have three close
snapshots, we notice that the deviation between the different radial
measurements is of order 10\% for $r_b$, 20\% for $r_{ch}$, and 5\%
for $r_{ck}$, but the deviation between the three different types of
core radii is larger.

The central surface brightness slope is obtained by calculating the
derivative of the smooth profile inside the core radius. This
derivative is constant for $r<r_b$. It is worth mentioning that the
value is the same when we measure the slope of a linear fit to the
photometric points in the same region. For the couple of very
concentrated cases, models m1t12.5 and m1t16.0, where we cannot
reliably measure a break radius, we take the central values of the
derivative as the slope.

If we try to use completeness uncorrected profiles instead of light
profiles the value for $r_h$, $r_b$ and $r_{ck}$ does not change much,
since it is the shape of the profile inside the core radius that
changes, but not the turnover radius. $r_{ch}$ on the other hand
suffers a larger change since the value of the central density is
lower. Obviously, the value of the central slope is very different
(flatter in general). If we try to construct a star count profile
using only those stars that are detected with close to 100\%
completeness, there are too few stars left and the profile becomes too
noisy in the center to make any meaningful measurements.

\begin{deluxetable}{lcccccc}
\tablewidth{0pt}
\tabletypesize{\huge}
\tablecaption{\label{tab2}Results.}
\tablehead{
\colhead{model} &
\colhead{SB slope}   &
\colhead{$r_{b}$} &
\colhead{$r_{ch}$}   &
\colhead{$r_{ck}$}   &
\colhead{$r_{h}$}   &\\
\colhead{} &
\colhead{}   &
\colhead{pc} &
\colhead{pc}   &
\colhead{pc}   &
\colhead{pc}   &
}
\startdata
m1t1.0  & -0.01 & 1.7 & 1.7 & 1.6 & 7.0 \\
m1t4.0  & ~0.07 & 1.6 & 1.6 & 1.6 & 5.2 \\
m1t7.0  & ~0.00 & 1.3 & 2.2 & 1.5 & 5.4 \\
m1t9.0  & -0.05 & 1.1 & 1.3 & 1.2 & 5.5 \\
m1t10.0  & -0.09 & 1.0 & 1.3 & 0.0 & 4.5 \\
m1t11.0  & -0.07 & 0.9 & 0.8 & 0.7 & 4.8 \\
m1t12.5  & -1.00 & \nodata & 0.3 & 0.4 & 4.1 \\
m1t16.0  & -0.71 & \nodata & 0.3 & 0.7 & 5.4 \\
m2t2.0  & ~0.04 & 5.4 & 5.1 & 5.5 & 10.8 \\
m2t6.0  & -0.05 & 4.5 & 4.5 & 3.8 & 7.2 \\
m2t8.0  & -0.05 & 2.6 & 2.0 & 1.7 & 5.8 \\
m3t2.0  & -0.07 & 2.9 & 2.7 & 4.0 & 9.0 \\
m3t5.0  &  0.18 & 2.2 & 4.1 & 3.4 & 8.9 \\
m3t8.0  & -0.11 & 2.6 & 2.4 & 3.3 & 8.3 \\
m3t11.0  & -0.08 & 2.2 & 2.2 & 2.9 & 8.4 \\
m3t14.0  &  0.08 & 1.8 & 2.3 & 2.0 & 7.4 \\
 & & & & & \\
\hline
 & & & & & \\
mb1t11.5 & -0.18 & 1.6 & 1.3 & 1.7 & 8.4 \\
mb2t11.0 & -0.26 & 1.8 & 0.9 & 1.6 & 8.8 \\
mb2t11.5 & -0.17 & 1.8 & 1.4 & 1.8 & 8.8 \\
mb2t12.0 & -0.18 & 1.9 & 1.3 & 1.8 & 8.7 \\
mb3t11.5 & -0.13 & 2.3 & 1.5 & 2.5 & 10.3 \\
mb4t3.0 & -0.20 & 2.3 & 0.7 & 1.4 & 7.0 \\
mb4t6.0 & -0.45 & 2.0 & 0.7 & 1.8 & 8.3 \\
mb4t9.0 & -0.07 & 2.0 & 1.8 & 2.2 & 8.7 \\
mb4t11.0 & -0.28 & 1.9 & 0.9 & 2.2 & 10.4 \\
mb4t11.5 & -0.07 & 2.6 & 1.7 & 2.2 & 9.6 \\
mb4t12.0 & -0.19 & 1.7 & 1.5 & 2.2 & 10.1 \\
mb5t11.3 & -0.17 & 1.8 & 1.3 & 1.7 & 8.8 \\
mb5t11.8 & -0.16 & 2.0 & 1.1 & 1.6 & 8.6 \\
mb5t12.0 & -0.39 & 2.1 & 0.7 & 1.4 & 8.4 \\
\enddata
\end{deluxetable}
\vspace{0cm}

Fig. 5 shows the fit of single-mass King profiles to model m3t, which
contains stellar-mass black holes. It can be seen that the fits are
significantly more noisy in the center than those for models m1t and
m2t, which is most likely due to the more stochastic heating of the
cluster by a few black holes as compared to a core of neutron stars
and white dwarfs. Table~2 shows that the derived photometric
parameters are nevertheless still within the range seen for those of
clusters without stellar-mass black holes, in particular the central
surface brightness slopes are still all below -0.12.

Once we have central surface brightness slopes and $r_c/r_h$
measurements for every modeled cluster we proceed to plot each point
on a slope versus $r_c/r_h$ plane. We create a plane for each of the
three measured radii. We find that using $r_{ck}$ or $r_{b}$ gives
very similar results, while using $r_{rh}$ does not give meaningful
constraints for clusters without flat central light profiles, so we
exclude this quantity from further analysis. In Figures~6 and~7 we
show the location of our models on the slope versus $r_b/r_h$ and
$r_{ck}/r_h$ planes respectively. The models span a range of central
slopes from 0.18 to -1.00, but the only models that have slopes
steeper than -0.5 are those that have achieved core collapse, while
the steepest slope for a model containing an IMBH is -0.45. As shown
in Figs~6 and 7, there are two models containing an IMBH that present
a flat central core (model mb4t9.0 and mb4t11.5), otherwise, only the
models containing intermediate-mass black holes show shallow central
cusps. Models without IMBHs show either a flat central slope, or steep
central cusps. Regarding the $r_b/r_h$ ratio, the cases that haven't
reached core-collapse and started from a King model with $W_0=7$ lie
within a narrow range between 0.15 and 0.35 and there is no clear
distinction between these cases and those containing IMBHs in this
respect. The cases that clearly separate towards large $r_c/r_h$ are
those that started from King models with $W_0=5$. The models
containing stellar-mass black holes lie close to the first group, but
have larger $r_b/r_h$ values. The two core-collapsed cases are placed
at $r_b/r_h=0$, since we cannot formally measure a break radius for
them. For the $r_{ck}/r_h$ case, the actual values change, but the
behavior is similar. The only group of models that clearly separates
from the rest in both plots are those with very steep central slopes
and non-detectable turnover radius, which correspond to clusters that
do not contain IMBHs and have undergone core-collapse.

We overlay on both planes all the Galactic clusters in NG06, plus
omega Centauri and G1. For G1 we measure the central slope using the
profile in \citet{geb05}, while $r_c$ and $r_h$ values come from the
analysis of \citet{ma07}. The first thing to notice is that the
Galactic clusters occupy a larger area in the plane than the modeled
ones. The two clusters for which there are kinematical indications of
hosting an IMBH, G1 and omega Centauri have central density slopes
shallower than -0.1, and their $r_c/r_h$ values are different. Omega
Centauri and G1 sit near the locus of our models, but both of them
have more extreme values of $r_c/r_h$ than the models with IMBHs. Very
concentrated clusters, like M15, which are assumed to have undergone
core-collapse, do lie very close to the models without IMBHs and long
evolutionary times. It should be noticed that some individual Galactic
clusters change location from one plane to the other

\section{Discussion and Conclusions}\label{analysis}

M15 was the first cluster for which the presence of a central massive
black hole was kinematically investigated, mainly due to its
concentrated central profile. It was only later when it became clear
that a projected steep central cusp is not the expected behavior for a
star cluster containing a black hole. This stresses the need to
develop better diagnostics to discriminate suitable candidates for
detailed kinematical measurements when looking for IMBHs. In this
paper we have created realistic synthetic images from N-body models of
star clusters with and without intermediate-mass black holes. We have
analyzed these images in the same way we analyze \textit{HST} data for
a sample of Galactic globular clusters and we compare both
datasets. We explore two quantities as possible diagnostic tools for
the presence of black holes: the central logarithmic slope of the
surface brightness profile, and the ratio of core radius to half light
radius. We find that the $r_c/r_h$ ratio cannot discriminate between
models with and without black holes, as \citet{hur07} already found,
but that the central logarithmic slope can. N-body clusters without
IMBHs show either flat central cores, or steep cusps if they have
undergone core-collapse, while clusters containing IMBHs show shallow
central cusp for all except two cases.

\begin{figure}[t]
  \centerline{\psfig{file=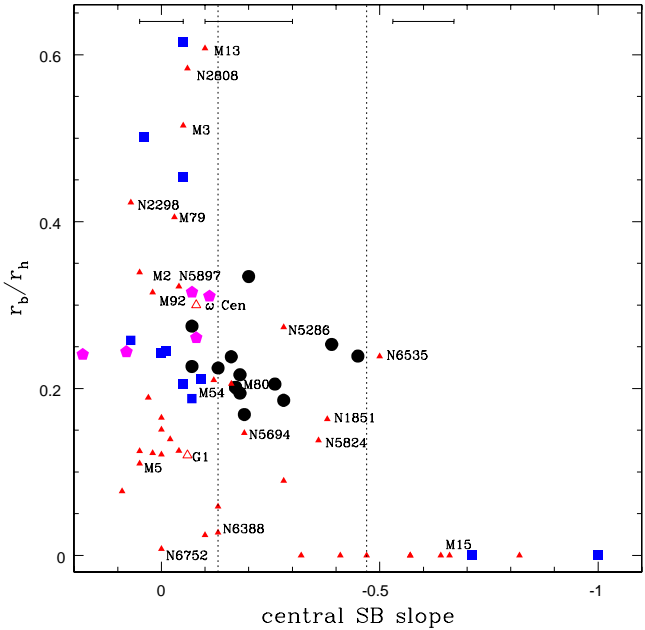,width=9.0cm,angle=0}}
  \figcaption{Central surface brightness slope versus the ratio of
    $r_b/r_h$.  The full circles mark the location of models
    containing IMBHs, the full squares are for models without an IMBH,
    and the full pentagons are for models containing stellar-mass
    black holes. Representative error bars for the central slopes
    (NG06) are shown on the top. The full triangles mark the location
    of 38 Galactic globular clusters, while the open triangles are for
    G1 and omega Cen. Some individual globular clusters are labeled.}
\label{radslope}
\end{figure}

\begin{figure}[t]
  \centerline{\psfig{file=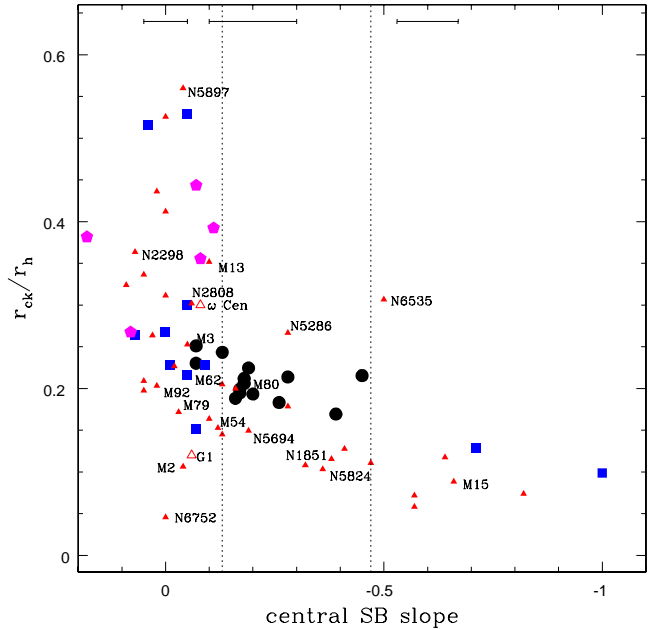,width=9.0cm,angle=0}}
  \figcaption{Central surface brightness slope versus the ratio of
    $r_{ck}/r_h$.  The filled circles mark the location of models
    containing IMBHs, the filled squares are for the models without an
    IMBH, and the full pentagons are for models containing
    stellar-mass black hole. As in the previous figure, representative
    error bars for the central slopes (NG06) are shown on the top, the
    full triangles mark the location of 38 Galactic globular clusters,
    while the open triangles are for G1 and omega Cen. Some individual
    globular clusters are labeled}
\label{radslope2}
\end{figure}

We want to emphasize that when dealing with density profiles of star
clusters, saying 'core radius' alone is not enough, one has to specify
how that radius is measured in order to compare its value to models or
other observations. Historically, the definition that we call $r_{ch}$
is the most popular one, but we show here that this definition is only
useful for clusters whose profiles have a flat central core. When the
profiles have central slopes different from zero, $r_{ck}$ or $r_b$
appear to be more suitable, although they can differ by up to a factor
of two for the same cluster. For density profiles with a flat central
core and clear turnovers, all three definitions mark practically the
same radius.

There are various ways in which our N-body simulations are idealized
compared to Galactic globular clusters. First, the number of stars and
the central densities are lower than for real clusters. Increasing
both quantities would increase the relaxation time, which in turn
would increase the evolutionary times. Including the presence of
binaries could change not only the timescales but also the nature of
the core contraction and expansion. Also, our analysis comes from
images with a limited amount of signal to noise (a combination of
number of stars and fiducial distance). It is likely that the
comparisons would be more meaningful using images with a larger
number of stars. These issues have to be kept in mind when comparing
to observed globular clusters. Despite the idealizations in the
models, the span of $r_c/r_h$ and central slope values seems to
generally agree between our models and observed clusters. We note that
the agreement between the two models that have undergone core-collapse
and the observed clusters that are suspect of having undergone the
same process is very good. Having simulations with a larger number of
stars would allow to analyze snapshots closer in time to fully explore
the process of core-collapse. There are some areas of Figs~6 and~7
containing observed clusters that our models do not populate. As
mentioned in section~\ref{intro}, a variety of heating mechanisms have
been proposed for star clusters in recent years. Some of them, like
mass loss or white dwarf kicks should affect most clusters; while
others like tidal shocking or primordial binaries depend on the
structure and evolution history of each cluster. A combination of
including some of these heating mechanisms and starting from a larger
variety of configurations (a larger range of initial $W_0$) would
likely produce a better agreement between models and observations.

Our results are in contrast with those of \citet{ves10} since they
find a number of models that present central shallow cusps without
containing black holes. We think the reason for the difference between
their result and ours lies on a combination of two things: on one
hand, their models contain about 10\% of the number of stars our
models have. On the other hand, they count main sequence stars, which
we find to be detected with a large degree of incompleteness in
realistic analysis, particularly in the center of rich clusters. Thus,
we are tracing a different subset of stars when measuring density
profiles. We believe that the lower numbers of stars in their models
produces noisier profiles that in turn can show shallow cusps due to
fluctuations in the photometric points. This is illustrated by the
fact that as soon as they use more particles (64K runs with combined
snapshots), their central slopes before core-collapse times converge
to shallower values consistent with the ones we find for models
without black holes.

It is clear from figures~6 and~7 that a division can be made between
clusters with and without black holes using only one of the two
quantities that we have explored, the central logarithmic slope of the
surface brightness profile. The $r_c/r_h$ ratio cannot distinguish
between cases with and without black holes, only between clusters that
have undergone core-collapse and the rest. Clusters that have achieved
core-collapse separate cleanly from the rest in both indicators, which
leads to the exclusion of very concentrated clusters, like M15, as
candidates for hosting an IMBH. None of our models reproduce central
slopes between 0.5 and 0.65 and we observe a few Galactic globular
clusters with those slopes. Since we are not able to follow the
details of the evolution right before core-collapse due to the time
intervals between snapshots, we cannot rule out the possibility that
clusters in this stage could have intermediate slopes Even if clusters
undergo such a phase, it is expected to be only for a very short
time. Two of the 14 models containing IMBHs do not show a clear
central shallow cusp. Even though it is impossible to draw statistical
conclusions from such a small sample, we can say that the absence of
shallow cusp does not imply the absence of a central black
hole. Therefore, some clusters with shallow cores might still be
interesting candidates to follow up with kinematics. Finally, clusters
with central slopes between -0.1 and -0.45 are clear candidates for
harboring central black holes since we can only reproduce shallow
central slopes by including intermediate-mass black holes. We conclude
that the central logarithmic surface brightness slope appears to be a
good diagnostic tool for choosing star clusters candidates for
harboring intermediate-mass black holes.

\acknowledgments

H.B. acknowledges support from the German Science foundation through a
Heisenberg Fellowship and from the Australian Research Council through
Future Fellowship grant FT0991052. The authors want to thank the
hospitality of the Kavli Institute for Theoretical Physics at UCSB, as
well as the organizers of the 'Formation and Evolution of Globular
Clusters' program there. We also thank the anonymous referee for
making useful suggestions that helped improve the manuscript.


\begin{thebibliography}{}

\bibitem[\protect\citeauthoryear{{Aarseth}}{{Aarseth}}{1999}]{aar99}
{Aarseth}, S.~J. 1999, \pasp, 111, 1333

\bibitem[{{Anderson} \& {van der Marel}(2010)}]{and10}
{Anderson}, J., \& {van der Marel}, R.~P. 2010, \apj, 710, 1032

\bibitem[\protect\citeauthoryear{{Bahcall} \& {Wolf}}{{Bahcall} \&
  {Wolf}}{1976}]{bah76}
{Bahcall}, J.~N.,  \& {Wolf}, R.~A. 1976, ApJ, 209, 214

\bibitem[\protect\citeauthoryear{{Bastian} et~al.}{{Bastian}
  et~al.}{2008}]{bas08}
{Bastian}, N., {Gieles}, M., {Goodwin}, S.~P., {Trancho}, G., {Smith}, L.~J.,
  {Konstantopoulos}, I.,  \& {Efremov}, Y. 2008, \mnras, 389, 223

\bibitem[\protect\citeauthoryear{{Baumgardt} et~al.}{{Baumgardt}
  et~al.}{2003a}]{bau03b}
{Baumgardt}, H., {Heggie}, D.~C., {Hut}, P.,  \& {Makino}, J. 2003a, MNRAS,
  341, 247

\bibitem[\protect\citeauthoryear{{Baumgardt} et~al.}{{Baumgardt}
  et~al.}{2003b}]{bau03a}
{Baumgardt}, H., {Hut}, P., {Makino}, J., {McMillan}, S.,  \& {Portegies
  Zwart}, S. 2003b, ApJL, 582, L21

\bibitem[\protect\citeauthoryear{{Baumgardt} \& {Kroupa}}{{Baumgardt} \&
  {Kroupa}}{2007}]{bau07}
{Baumgardt}, H.,  \& {Kroupa}, P. 2007, \mnras, 380, 1589

\bibitem[\protect\citeauthoryear{{Baumgardt}, {Makino}, \&
  {Ebisuzaki}}{{Baumgardt} et~al.}{2004a}]{bau04a}
{Baumgardt}, H., {Makino}, J.,  \& {Ebisuzaki}, T. 2004a, \apj, 613, 1133

\bibitem[\protect\citeauthoryear{{Baumgardt}, {Makino}, \&
  {Ebisuzaki}}{{Baumgardt} et~al.}{2004b}]{bau04b}
{Baumgardt}, H., {Makino}, J.,  \& {Ebisuzaki}, T. 2004b, \apj, 613, 1143

\bibitem[\protect\citeauthoryear{{Baumgardt}, {Makino}, \& {Hut}}{{Baumgardt}
  et~al.}{2005}]{bau05}
{Baumgardt}, H., {Makino}, J.,  \& {Hut}, P. 2005, ApJ, 620, 238

\bibitem[\protect\citeauthoryear{{Baumgardt} et~al.}{{Baumgardt}
  et~al.}{2003c}]{bau03c}
{Baumgardt}, H., {Makino}, J., {Hut}, P., {McMillan}, S.,  \& {Portegies
  Zwart}, S. 2003c, ApJL, 589, L25

\bibitem[\protect\citeauthoryear{{Beers}, {Flynn}, \& {Gebhardt}}{{Beers}
  et~al.}{1990}]{bee90}
{Beers}, T.~C., {Flynn}, K.,  \& {Gebhardt}, K. 1990, AJ, 100, 32

\bibitem[{{Casertano} \& {Hut}(1985)}]{cas85}
{Casertano}, S., \& {Hut}, P. 1985, \apj, 298, 80

\bibitem[\protect\citeauthoryear{{Cohn}}{{Cohn}}{1980}]{coh80}
{Cohn}, H. 1980, ApJ, 242, 765

\bibitem[\protect\citeauthoryear{{Davis} et~al.}{{Davis} et~al.}{2008}]{dav08}
{Davis}, D.~S., {Richer}, H.~B., {King}, I.~R., {Anderson}, J., {Coffey}, J.,
  {Fahlman}, G.~G., {Hurley}, J.,  \& {Kalirai}, J.~S. 2008, \mnras, 383, L20

\bibitem[\protect\citeauthoryear{{Drukier}}{{Drukier}}{1995}]{dru95}
{Drukier}, G.~A. 1995, \apjs, 100, 347

\bibitem[\protect\citeauthoryear{{Drukier} et~al.}{{Drukier}
  et~al.}{2003}]{dru03}
{Drukier}, G.~A., {Bailyn}, C.~D., {Van Altena}, W.~F.,  \& {Girard}, T.~M.
  2003, \aj, 125, 2559

\bibitem[\protect\citeauthoryear{{Drukier}, {Fahlman}, \& {Richer}}{{Drukier}
  et~al.}{1992}]{dru92}
{Drukier}, G.~A., {Fahlman}, G.~G.,  \& {Richer}, H.~B. 1992, \apj, 386, 106

\bibitem[\protect\citeauthoryear{{Dull} et~al.}{{Dull} et~al.}{1997}]{dul97}
{Dull}, J.~D., {Cohn}, H.~N., {Lugger}, P.~M., {Murphy}, B.~W., {Seitzer},
  P.~O., {Callanan}, P.~J., {Rutten}, R.~G.~M.,  \& {Charles}, P.~A. 1997, ApJ,
  481, 267

\bibitem[\protect\citeauthoryear{{Ferraro} et~al.}{{Ferraro}
  et~al.}{2003}]{fer03}
{Ferraro}, F.~R., {Possenti}, A., {Sabbi}, E., {Lagani}, P., {Rood}, R.~T.,
  {D'Amico}, N.,  \& {Origlia}, L. 2003, ApJ, 595, 179

\bibitem[\protect\citeauthoryear{{Fregeau}}{{Fregeau}}{2008}]{fre08}
{Fregeau}, J.~M. 2008, \apjl, 673, L25

\bibitem[\protect\citeauthoryear{Fregeau et~al.}{Fregeau et~al.}{2009}]{fre09}
Fregeau, J.~M., Richer, H.~B., Rasio, F.~A.,  \& Hurley, J.~R. 2009, The
  dynamical effects of white dwarf birth kicks in globular star clusters

\bibitem[\protect\citeauthoryear{{Gao} et~al.}{{Gao} et~al.}{1991}]{gao91}
{Gao}, B., {Goodman}, J., {Cohn}, H.,  \& {Murphy}, B. 1991, ApJ, 370, 567

\bibitem[\protect\citeauthoryear{{Gebhardt}, {Rich}, \& {Ho}}{{Gebhardt}
  et~al.}{2005}]{geb05}
{Gebhardt}, K., {Rich}, R.~M.,  \& {Ho}, L.~C. 2005, ApJ, 634, 1093

\bibitem[\protect\citeauthoryear{{Gebhardt} et~al.}{{Gebhardt}
  et~al.}{2003}]{geb03}
{Gebhardt}, K., et~al. 2003, \apj, 583, 92

\bibitem[\protect\citeauthoryear{{Gerssen} et~al.}{{Gerssen}
  et~al.}{2002}]{ger02}
{Gerssen}, J., {van der Marel}, R.~P., {Gebhardt}, K., {Guhathakurta}, P.,
  {Peterson}, R.~C.,  \& {Pryor}, C. 2002, AJ, 124, 3270

\bibitem[\protect\citeauthoryear{{Gerssen} et~al.}{{Gerssen}
  et~al.}{2003}]{ger03}
{Gerssen}, J., {van der Marel}, R.~P., {Gebhardt}, K., {Guhathakurta}, P.,
  {Peterson}, R.~C.,  \& {Pryor}, C. 2003, AJ, 125, 376

\bibitem[\protect\citeauthoryear{{Harris}}{{Harris}}{1996}]{har96}
{Harris}, W.~E. 1996, AJ, 112, 1487

\bibitem[\protect\citeauthoryear{{Hurley}}{{Hurley}}{2007}]{hur07}
{Hurley}, J.~R. 2007, \mnras, 379, 93

\bibitem[\protect\citeauthoryear{{Hurley}, {Pols}, \& {Tout}}{{Hurley}
  et~al.}{2000}]{hur00}
{Hurley}, J.~R., {Pols}, O.~R.,  \& {Tout}, C.~A. 2000, \mnras, 315, 543

\bibitem[\protect\citeauthoryear{{Irwin} et~al.}{{Irwin}
  et~al.}{2010}]{irw10}
{Irwin}, J.~A. and {Brink}, T.~G. and {Bregman}, J.~N. and {Roberts}, T.~P. 2010, ApJL, 712, L1

\bibitem[\protect\citeauthoryear{{King}}{{King}}{1962}]{kin62}
{King}, I. 1962, \aj, 67, 471

\bibitem[\protect\citeauthoryear{{King}}{{King}}{1966}]{kin66}
{King}, I.~R. 1966, AJ, 71, 276

\bibitem[\protect\citeauthoryear{{Kochanek}}{{Kochanek}}{1992}]{koc92}
{Kochanek}, C. 1992, AJ, 385, 604


\bibitem[\protect\citeauthoryear{{Kroupa}}{{Kroupa}}{2001}]{kro01}
{Kroupa}, P. 2001, \mnras, 322, 231

\bibitem[\protect\citeauthoryear{{Lanzoni} et~al.}{{Lanzoni}
  et~al.}{2007}]{lan07}
{Lanzoni}, B., {Dalessandro}, E., {Ferraro}, F., {Miocchi}, P., {Valenti}, E.,  \& {Rood}, R.~T. 2007, \apjl, 668, L139


\bibitem[\protect\citeauthoryear{{Lugger}, {Cohn}, \& {Grindlay}}{{Lugger}
  et~al.}{1995}]{lug95}
{Lugger}, P.~M., {Cohn}, H.~N.,  \& {Grindlay}, J.~E. 1995, ApJ, 439, 191

\bibitem[\protect\citeauthoryear{{L\"utzgendorf} et~al.}{{L\"utzgendorf} et~al.}{2011}]{lut11}
{L\"utzgendorf}, N., et~al. 2011, arxiv, 1107.4243

\bibitem[\protect\citeauthoryear{{Ma} et~al.}{{Ma} et~al.}{2007}]{ma07}
{Ma}, J., et~al. 2007, \mnras, 376, 1621

\bibitem[\protect\citeauthoryear{{Maccarone} et~al.}{{Maccarone}
  et~al.}{2007}]{mac07}
{Maccarone}, T.~J., {Kundu}, A., {Zepf}, S.~E.,  \& {Rhode}, K.~L. 2007, \nat,
  445, 183

\bibitem[\protect\citeauthoryear{{Mackey} et~al.}{{Mackey}
  et~al.}{2008}]{mac08}
{Mackey}, A.~D., {Wilkinson}, M.~I., {Davies}, M.~B.,  \& {Gilmore}, G.~F.
  2008, \mnras, 386, 65

\bibitem[\protect\citeauthoryear{{Magorrian} et~al.}{{Magorrian}
  et~al.}{1998}]{mag98}
{Magorrian}, J., et~al. 1998, \aj, 115, 2285

\bibitem[\protect\citeauthoryear{{Matsumoto} \& {Tsuru}}{{Matsumoto} \&
  {Tsuru}}{1999}]{mat99}
{Matsumoto}, H.,  \& {Tsuru}, T.~G. 1999, \pasj, 51, 321

\bibitem[\protect\citeauthoryear{{Matsumoto} et~al.}{{Matsumoto}
  et~al.}{2001}]{mat01}
{Matsumoto}, H., {Tsuru}, T.~G., {Koyama}, K., {Awaki}, H., {Canizares}, C.~R.,
  {Kawai}, N., {Matsushita}, S.,  \& {Kawabe}, R. 2001, \apjl, 547, L25

\bibitem[\protect\citeauthoryear{{McCrady}, {Gilbert}, \& {Graham}}{{McCrady}
  et~al.}{2003}]{mcc03}
{McCrady}, N., {Gilbert}, A.~M.,  \& {Graham}, J.~R. 2003, \apj, 596, 240

\bibitem[\protect\citeauthoryear{{McNamara}, {Harrison}, \&
  {Baumgardt}}{{McNamara} et~al.}{2004}]{mcn04}
{McNamara}, B.~J., {Harrison}, T.~E.,  \& {Baumgardt}, H. 2004, \apj, 602, 264

\bibitem[\protect\citeauthoryear{{Noyola} \& {Gebhardt}}{{Noyola} \&
  {Gebhardt}}{2006}]{noy06}
{Noyola}, E.,  \& {Gebhardt}, K. 2006, \aj, 132, 447

\bibitem[\protect\citeauthoryear{{Noyola} \& {Gebhardt}}{{Noyola} \&
  {Gebhardt}}{2007}]{noy07}
{Noyola}, E.,  \& {Gebhardt}, K. 2007, \aj, 134, 912

\bibitem[\protect\citeauthoryear{{Noyola}, {Gebhardt}, \& {Bergmann}}{{Noyola}
  et~al.}{2008}]{noy08}
{Noyola}, E., {Gebhardt}, K.,  \& {Bergmann}, M. 2008, \apj, 676, 1008

\bibitem[\protect\citeauthoryear{{Pfahl}, {Rappaport}, \&
    {Podsiadlowski}}{{Pfahl} et~al.}{2002}]{pfa02a} {Pfahl}, E.,
  {Rappaport}, S., \& {Podsiadlowski}, P. 2002, \apj, 573, 283

\bibitem[\protect\citeauthoryear{{Pfahl} et~al.}{{Pfahl}
    et~al.}{2002}]{pfa02b} {Pfahl}, E., {Rappaport}, S.,
  {Podsiadlowski}, P., \& {Spruit}, H. 2002, \apj, 574, 364

\bibitem[\protect\citeauthoryear{{Pooley} \& {Rappaport}}{{Pooley} \&
  {Rappaport}}{2006}]{poo06}
{Pooley}, D.,  \& {Rappaport}, S. 2006, \apjl, 644, L45

\bibitem[\protect\citeauthoryear{{Stetson}}{{Stetson}}{1987}]{ste87}
{Stetson}, P.~B. 1987, PASP, 99, 191

\bibitem[\protect\citeauthoryear{{Strohmayer} \& {Mushotzky}}{{Strohmayer} \&
  {Mushotzky}}{2003}]{str03}
{Strohmayer}, T.~E.,  \& {Mushotzky}, R.~F. 2003, \apjl, 586, L61

\bibitem[\protect\citeauthoryear{{Trager}, {King}, \& {Djorgovski}}{{Trager}
  et~al.}{1995}]{tra95}
{Trager}, S.~C., {King}, I.~R.,  \& {Djorgovski}, S. 1995, AJ, 109, 218

\bibitem[\protect\citeauthoryear{{Trenti} et~al.}{{Trenti}
  et~al.}{2007}]{tre07}
{Trenti}, M., {Ardi}, E., {Mineshige}, S.,  \& {Hut}, P. 2007, \mnras, 374, 857

\bibitem[\protect\citeauthoryear{{Ulvestad}, {Greene}, \& {Ho}}{{Ulvestad}
  et~al.}{2007}]{ulv07}
{Ulvestad}, J.~S., {Greene}, J.,  \& {Ho}, L. 2007, ApJL, accepted

\bibitem[\protect\citeauthoryear{{van den Bosch} et~al.}{{van den Bosch}
  et~al.}{2006}]{bos06}
{van den Bosch}, R., {de Zeeuw}, T., {Gebhardt}, K., {Noyola}, E.,  \& {van de
  Ven}, G. 2006, ApJ, 641, 852

\bibitem[{{van der Marel} \& {Anderson}(2010)}]{vdm10}
{van der Marel}, R.~P., \& {Anderson}, J. 2010, \apj, 710, 1063

\bibitem[\protect\citeauthoryear{{Vesperini} \& {Chernoff}}{{Vesperini} \&
  {Chernoff}}{1994}]{ves94}
{Vesperini}, E.,  \& {Chernoff}, D.~F. 1994, \apj, 431, 231

\bibitem[\protect\citeauthoryear{{Vesperini} \& {Trenti}}{{Vesperini} \&
  {Trenti}}{2010}]{ves10}
{Vesperini}, E.,  \& {Trenti}, M. 2010, \apjl, 720, 179

\bibitem[\protect\citeauthoryear{Wahba \& Wang}{Wahba \& Wang}{1990}]{wah90}
Wahba, G.,  \& Wang, Y. 1990, Communications in Statistics, Part A -- Theory
  and Methods, 19, 1685

\bibitem[\protect\citeauthoryear{{Zepf} et~al.}{{Zepf} et~al.}{2008}]{zep08}
{Zepf}, S.~E., et~al. 2008, \apjl, 683, L139

\end{thebibliography}
\end{document}